\documentclass[11pt]{article}

\usepackage{amsthm}
\usepackage{amsmath}
\usepackage[boxed]{algorithm}
\usepackage{algorithmic}
\usepackage{amsmath, amsfonts}
\usepackage{amscd}
\usepackage{amssymb}
\usepackage{amsxtra}
\newcommand{\calO}{{\cal{O}}}

\newcommand{\ldotdot}{\,.\,.\,}

\newtheorem{thm}{Theorem}[section]
\newtheorem{lem}[thm]{Lemma}
\newtheorem{claim}{Claim}[section]

\newtheorem{rem}[thm]{Remark}

\newtheorem{dfn}[thm]{Definition}

\newcommand{\be}{\begin{equation}}
\newcommand{\ee}{\end{equation}}
\newcommand{\bd}{\begin{displaymath}}
\newcommand{\ed}{\end{displaymath}}

\long\def\symbolfootnote[#1]#2{\begingroup%
\def\thefootnote{\fnsymbol{footnote}}\footnote[#1]{#2}\endgroup}

\let\abs=\envert

\newcommand{\bfE}{\textbf{E}}
\newcommand{\bfe}{\textbf{e}}

\newcommand{\dist}{\textbf{d}}
\newcommand{\Dist}{\textbf{Dist}}

\newcommand{\junk}[1]{}

\begin{document}
\title{Approximating General Metric Distances Between a Pattern and a Text}

%
%
%

\author{
Ely Porat\thanks{Department of Computer Science, Bar-Ilan
University, 52900 Ramat-Gan, Israel; {\tt porately@cs.biu.ac.il}}
\and Klim Efremenko\thanks{Department of Computer Science, Bar-Ilan
University, 52900 Ramat-Gan, Israel } \thanks{ Weizmann Institute,
Rehovot 76100, Israel}}


\maketitle

\begin{abstract}
Let $T=t_0 \ldots t_{n-1}$ be a text and $P = p_0 \ldots p_{m-1}$
a pattern taken from some finite alphabet set $\Sigma$, and let
$\dist$ be a metric on $\Sigma$. We consider the problem of
calculating the sum of distances between the symbols of $P$ and
the symbols of substrings of $T$ of length $m$ for all possible
offsets. We present an $\varepsilon$-approximation algorithm for
this problem which runs in time $O(\frac{1}{\varepsilon^2}n\cdot
\mathrm{ polylog}(n,\abs{\Sigma}))$.





\end{abstract}

\setlength{\parindent}{0.0in} \setlength{\parskip}{0.1 in}
\section{Introduction \label{sec:intro}}
\emph{String matching}, the problem of finding all occurrences of a given
pattern in a given text, is a classical problem in computer
science. The problem has pleasing theoretical features and a
number of direct applications to ``real world'' problems.

 Advances in multimedia, digital libraries, and computational
biology, have shown that a much more generalized theoretical basis
of string matching could be of tremendous
benefit~\cite{Pentland-92,olson:95}.
 To this end, string matching has had to adapt itself to
increasingly broader definitions of ``matching''. Two types of
problems need to be addressed -- {\it generalized matching} and
{\it approximate matching}. In generalized matching, one seeks all
exact occurrences of the pattern in the text, but the ``matching''
relation is defined differently. The output is all locations in
the text where the pattern ``matches'' according to the new
definition of a match. The different applications define the
matching relation. Examples can be seen in Baker's {\it
parameterized matching}~(\cite{bak:93}) or Amir and Farach's {\it
less-than matching}~(\cite{AF-95}). The second model, and the one
we are concerned with in this paper, is that of {\em approximate
matching}. In approximate matching, one defines a distance metric
between the objects (e.g.\ strings, matrices) and seeks to
calculate this distance for all text locations. Usually we seek
locations where this distance is small enough.

One of the earliest and most natural metrics is the {\it Hamming
distance}, where the distance between two strings is the number of
mismatching characters exists algorithm calculating this distance
exactly ~\cite{ALP0:00} and approximating it ~\cite{karloff:93}.
Levenshtein~\cite{L-66} identified three types of errors:
mismatches, insertions, and deletions. These operations are
traditionally used to define the {\it edit distance} between two
strings. The edit distance is the {\it minimum} number of edit
operations one needs to perform on the pattern in order to achieve
an exact match at the given text location. Lowrance and
Wagner~\cite{lw-75,wagner-75} added the {\it swap} operation to
the set of operations defining the distance metric.
 Much of the recent research in string matching
concerns itself with understanding the inherent ``hardness'' of
the various distance metrics, by seeking upper and lower bounds
for string matching under these conditions.

A natural subset of these problems is when the distance is defined
only on the alphabet, and therefore, the distance between two
strings is the sum of the distances between the corresponding characters
in both strings. It is possible to solve this problem in time
$O(n\cdot m)$ by employing the naive approach of summing the
distances between each character of the pattern, and its
corresponding character in the text, for each possible alignment
of the pattern. This problem was first defined by Muthukrishnan
in~\cite{muthu-open} and has been open since. In this paper we
present an approximation algorithm for this problem.

This algorithm
 consists of two parts: the first part is a preprocessing phase in which
random hash functions on the alphabet is constructed. We use same
hashing which Bartal used for tree embedding at ~\cite{Bar96}. We
use this hashing in order to separate the places where distance
between letters is large and places where this distance is small.


 The second part of the algorithm is an application of sampling~(\cite{CPER07}),
which allows us to give an approximation of the distance between
the text and the pattern, in time $O(\frac{1}{\varepsilon^2}n\cdot
\mathrm{polylog}(n,\abs{\Sigma}))$.

The contributions of this paper are twofold: on the technical
side, we have solved a problem that has been {\em open} for over a
decade, by presenting the {\em fastest} known approximation
algorithm for many metrics; additionally, and this is perhaps the
more important contribution of this paper, we have identified and
exploited a new technique -- {\it sampling}, that has been used in
some recent papers~(\cite{CPER07}) only implicitly. We employ
sampling in a much more sophisticated manner and show how to use
this important tool for approximating distances. We also present a
novel way of using embeddings and geometry tools in pattern
matching. This technique possesses a wide range of applications.
For example, one can easily extend it to calculate the
$\ell_2$-norm distance (in other words, when
\begin{equation}
\Dist(T[i \,.\,.\, i+m-1],P)=\sqrt{\sum_{j=0}^{m-1}
\dist(t_{i+j},p_j)^2}
\end{equation}
is the distance measure), or it can be extended for many infinite
metrics. Our algorithm also allows for symbols in a text to be
wildcards.

We believe that this new method for solving approximate string matching
problems -- embedding metric in some suitable space and sampling
-- may actually yield efficient algorithms for many more problems in the future.

\section{Problem definition and Preliminaries }\label{s:def}
\begin{dfn} A {metric space}
 is a pair $(X,\dist)$ where $X$ is a set of points and $ \dist:X
\times X\rightarrow {\bf{R^+}}$ is a {\em metric} satisfying the
following axioms:
\begin{equation}
\begin{split}
&\dist(x_1,x_2)=\dist(x_2,x_1).\\
&\dist(x_1,x_3) \leq \dist(x_1,x_2)+\dist(x_2,x_3). \\
&\dist(x_1,x_2)=0 \Leftrightarrow x_1=x_2.
\end{split}
\end{equation}
\end{dfn}
Let $(\Sigma,\dist)$ be a metric space, and $A= a_0 \ldots
a_{m-1}$ ,$B= b_0 \ldots b_{m-1}$ be any two strings of the same
length with symbols from $\Sigma$. We define the \emph{distance}
of $A$ and $B$ by $\Dist(A,B)=\sum_{i=0}^{m-1}\dist(a_i,b_i) $.
Given a text $T=t_0 \ldots t_{n-1}$ and a pattern $P=p_0 \ldots
p_{m-1}$, our goal is to calculate the array $S[i]=\Dist(T[i
\ldotdot i+m-1],P)$, for each possible offset $i=0, \ldots,
n-m-1$. Calculating the exact values of $S$ can be done in $O(nm)$
time, using the naive approach, In most cases it is enough to know
only an approximation of the distance; we therefore present an
efficient algorithm which approximates the values of $S$.
%

Convolutions can be used in the standard fashion to improve the time
for finite fixed alphabets.
\begin{dfn}  \label{def:FFT}
Let $A[0],\ldots ,A[n-1]$ and $B[0],\ldots ,B[m-1]$ be arrays of
natural numbers.
The {\em discrete convolution (polynomial multiplication) of $A$ and
$B$} is $V$ where:
\begin{equation}
V[i] = \sum_{j=0}^{m-1} A[i-j] B[j]\ ,
\end{equation} where
$i=0,\ldots, n-m$. We denote $V$ as $A*B$. We will choose $A=T$ and
$B=P$, i.e.\ we will treat $T$ and $P$ as the coefficients of
polynomials of degrees $n-1$ and $m-1$, respectively.

By standard tricks, namely, (1) reversing the text to obtain $T^R$;
(2) calculating $T^R * P$; (3) reverse the result; (4) discard the
first $m-1$ values, and last $n-m+1$ values of the result, we obtain
an array $V = (T^R * P)^R[m-1 \ldotdot n-1]$ where for each $i$,
\begin{equation}
V[i] = \sum_{j=0}^{m-1} t_{i+j} p_j\ .
\end{equation}
In other words, for every possible offset $i$, $V[i]$ is the sum of
the pattern symbols multiplied, each with its corresponding text
symbol. For convenience, we define $T\otimes P = (T^R * P)^R[m-1
\ldotdot n-1]$
\end{dfn}
A convolution can be computed in time $O(n\log m)$, in a
computational model with word size $O(\log m)$, by using the Fast
Fourier Transform (FFT)~\cite{CLR-92}.
\begin{rem}
Using FFT we can compute general pattern distances, in time
$\calO(\abs{\Sigma} n\log m)$, by using the following method: for
every $a \in \Sigma$, we define an array $\chi_a(P)$ by setting
$\chi_a(P)[i]=\chi_a(P[i])$, where $\chi_a(x)=1$ if $x=a$ and $0$
otherwise. Set $T_a[i]=\dist(a,t_i)$. Computing $T_a \otimes
\chi_a(P)$ gives us the sum of the distances of the letter $a$ from
the text. The sum of all convolution results is the desired
distance.
\end{rem}

We provide some required definitions regarding metric spaces:
\begin{dfn}
Let $\rho$ be a mapping $\rho \colon X \rightarrow Y$, where
$(X,\dist_1)$ and $(Y,\dist_2)$ are metric spaces. $\rho$ is an
\emph{isometry} if $\forall x,y \in X\
\dist_1(x,y)=\dist_2(\rho(x),\rho(y))$.
\end{dfn}
Unfortunately, we cannot always construct an isometry. Therefore we
consider weaker conditions:
\begin{dfn}
\label{def:D-emb} Given two metric spaces $(X,\dist_1)$ and
$(Y,\dist_2)$, and a value $c \geq 1$, a mapping $\rho:X \rightarrow
Y$ is called a {\em c-distortion embedding}, if for all $x,y \in X$,
\be \dist_1(x,y)\leq \dist_2(\rho(x),\rho(y))\leq
c\cdot\dist_1(x,y).
 \ee
\end{dfn}

\section{The One-Mismatch Algorithm}
In this section we describe the \emph{one-mismatch} algorithm,
in itself a very useful general tool in pattern matching. The
one-mismatch algorithm had been described before in
~\cite{CC07,CPER07}.

Given a numeric text $T$ and a numeric pattern $P$, and we want to
find exact matches of $P$ in $T$. One way to do so is by
calculating for each location $0\leq i \leq n-m-1$ the value:
\[
    \sum_{j=0}^{m-1}(p_j-t_{i+j})^2=\sum_{j=0}^m(p_j^2-2p_jt_{i+j}+t_{i+j}^2)\
    .
\]
Notice this sum will be zero iff there is an exact match at location
$i$. Furthermore this sum can be computed efficiently for all $i$'s
in $\calO(n\log{m})$ time using convolutions.  Notice that if $P$
and $T$ are not numeric then an arbitrary one-to-one mapping can be
chosen from the alphabet to the set of positive integers
${{N}}$.

This method can be extended to the case of matching with ``don't
cares''~\cite{CC07}, by simply calculating instead
\[
    A_0[i]= \sum_{j=0}^{m-1} p_j't'_{i+j}(p_j-t_{i+j})^2\ ,
\]
where $p'_j=0$ (resp.\ $t'_j=0$) if $p_j$ (resp.\ $t_j$) is a ``don't
care'' symbol and $1$ otherwise. Wherever there is an exact match with
``don't cares'' this sum will be exactly $0$. This can also be computed
with convolutions in time $\calO(n\log{m})$.

Again, this scheme can be further extended to the one-mismatch
problem, which is to determine if $P$ matches $T[i\ldotdot i+m-1]$
with at most one mismatch. Furthermore, we can identify the location
of the mismatch for each such $i$. This is done by also computing,
for each $i$,
\[
    A_1[i]=\sum_{j=0}^{m-1} jp_j't'_{i+j}(p_j-t_{i+j})^2\ ,
\]
by using the convolution. Then if $p$ matches the text at offset
$i$ with one mismatch then eventually $A_0[i]=(p_r-t_{i+r})^2$ and
$A_1[i]=r(p_r-t_{i+r})^2$ where $r$ is a location of a mismatch.
Therefore, by calculating $A_1[i]/A_0[i]$, we find the supposed
mismatch location and verify it. Finally, locations where exact
matches occur will be labeled ``match'', location where a single
mismatch occurs will be labeled with its location, and location
where more than one mismatch occurs will be labeled $\bot$. The
one-mismatch algorithm is therefore as follows:
\begin{enumerate}
    \item compute the array     $$A_0[i]= \sum_{j=0}^m
    p_j't'_{i+j}(p_j-t_{i+j})^2$$ using FFT.
    \item  compute the array $$A_1[i]=\sum_{j=0}^m
    jp_j't'_{i+j}(p_j-t_{i+j})^2$$ using FFT.
    \item  If $A_0[i] = 0$ set $B[i]\gets\mbox{``match''}$ else set
    $B[i]\gets A_1[i]/A_0[i]$.
    \item For each $i$ s.t.\ $B[i]\neq \mbox{``match''}$, check to see if
    $(p[B[i]]-t[B[i]+i])^2=A_0[i]$. If this is not the case then
    set $B[i]\gets \bot$.
\end{enumerate}
The running time of this algorithm is $\calO(n\log{m})$.

\section{The Sampling Method}\label{s:naive}
In this section we present a general method referred to as the
\emph{sampling} method. It allows us, for every possible offset,
to sample (i.e.\ choose) a {\em random mismatch} from the set of all
mismatches w.r.t.\ this offset.
We show how to utilize the previously described algorithm for this
purpose.

First fix some probability $0< q \leq 1$, and define subpattern
$P^*$ of $P$ by:

\be \label{ri}
 p^{*}_i=
 \begin{cases}
 p_i  & \mbox{with probability } q;\\
\phi \mbox{(don't care)}             & \mbox{otherwise.}
 \end{cases}
 \ee

In the algorithm referred to as Sample($q,T,P$), we simply create
$P^*$ as defined in~\eqref{ri} and run the one-mismatch algorithm on
$P^*$ and $T$. Now, for every offset $i$, let $m_i$ be the number of
mismatches between $T$ and $P$ w.r.t.\ this offset. The following
lemma trivially follows:

\begin{lem}\label{sample}
Let $B$ be the array returned by Sample($q,T,P$). For some
location $i$,
\[ \Pr(B[i]=\mbox{``match''})=(1-q)^{m_i}\ ,\]
and
\[ \Pr(\mbox{$B[i]$ is a
mismatch location})=m_iq(1-q)^{m_i-1}\ .\]
\end{lem}

Another important property of this algorithm is that the mismatch
returned w.r.t.\ offset $i$ is uniformly distributed over the set
of all mismatches w.r.t.\ offset $i$. We will show how to use this
algorithm to sample a random location for which the distance is
not $0$. Notice that for $q\approx \frac{1}{m_i}$, the probability
of finding a mismatch w.r.t.\ offset $i$ is $\calO(1)$. Therefore,
we can enumerate on $q=\{ 2^{-j} \}_{j=0}^{\log{m}}$. Then, for
every location $i$ there exists some $q$ which is
$\approx\frac{1}{m_i}$. Therefore, the next algorithm finds for
each location a mismatch with constant probability, which is
uniformly distributed over all the mismatches.
\begin{enumerate}
\item {\bf{for}} $q=1$; $q\geq 1/m$ ; $q=q/2$ \item \ \ \
Sample(q,T,P) \item For every offset $i$ if a mismatch is found
return it.
\end{enumerate}

\section{Motivation for the Algorithm}\label{s:Motiv}
\begin{rem}
In this paper we assume that the ratio of maximal and minimal
distances is bounded by $B_{\dist}$. Therefore, w.l.o.g.\ we can
assume that the minimal nonzero distance is $1$, i.e.\ $\forall
x,y \in \Sigma,\ \dist(x,y)\leq B_{\dist}$ and $\dist(x,y)>0 \iff
\dist(x,y) \geq 1$. That is because if $B_{\min}$ is the minimal
distance, then we can use the metric $\frac{\dist}{B_{\min}}$
instead.
\end{rem}
A first naive approach to approximate the distance is as follows:
say we wish to provide an approximation only for some offset $i$,
and let $X$ be a random variable which is equal to
$\dist(t_{i+j},p_j)$, where $j$ is chosen uniformly from
$0,1,\ldots,m-1$. We can sample $X$ by choosing a random $j$ and
calculating $\dist(t_{i+j},p_j)$. The expectation of $m \cdot X$ is
the desired sum. Therefore, the way to compute $\bfE(X)$ is sample
$X$ several times and return the average. The problem with this
approach is that the variance of $X$ may be very large: for example,
if $P$ matches $T$ except for a few mismatches, then w.h.p.\ we will
not sample even a single mismatch.

The second attempt to reduce the variance of variable $X$, is to
use the sampling algorithm described in Sect.~\ref{s:naive}. As a
result, $X$ will be distributed only over locations where
$\dist(t_{i+j},p_j)>0$. That is because the sampling algorithm
returns only relative locations $j$ for which $t_{i+j} \neq p_j$.
This sampling approach reduces the variance of $X$, but still it
may happen that for some offset $i$, all distances are very small
except for a single one which is even greater than the sum of all
others. With high probability, only the smaller distances will be
sampled, thus affecting the final outcome. This approach can
provide us with an algorithm which runs in $O(n\cdot
 B_{\dist}\cdot \mathrm{polylog}(n,\abs{\Sigma}))$ time, however,
$B_{\dist}$ may be very large.

All the above leads us to search for a way to sample only
locations $j$ for which, when some $D$ is fixed, $D \leq
\dist(t_{i+j},p_j)<2D$. Then, with an additional multiplicative
factor of $\log(B_{\dist})$, we can enumerate on
$D=\{2^i\}_{i=0}^{\log{B_{\dist}}}$, each step approximating the
expectation of the variable $X_D$ which uniformly ranges over $\{
\dist(t_{i+j},p_j) \mid  D \leq \dist(t_{i+j},p_j) < 2D\}$. Notice
that for any value $a$,
\[
   \Pr(X_D=a)=
     \begin{cases}
     \frac{\Pr(X=a)}{\Pr(D\leq X<2D)} & D\leq a <2D\mbox{,} \\
     0              & \mbox{else}.
     \end{cases}
\]
Hence it follows that
$$\bfE(X)=\sum_D \Pr(D\leq X <2D)\bfE(X_D)\ .$$
A hypothetic way to sample $X_D$ would be to design a mapping
$\pi_D$ on $\Sigma$ for which
\begin{equation} \label{Mapping}
D<\dist(x,y)<2D \iff \pi_D(x) \neq \pi_D(y)\ .
\end{equation}
If so, we could have run the sampling algorithm on
$\pi_{\dist}(T)$ and $\pi_{\dist}(P)$, applying $\pi_{\dist}$ in
the obvious way, and obtain samples of $X_D$. Then, the average of
sampled distances will approximate $\bfE(X_D)$. The approximation
of $\Pr(D\leq X <2D)$ would have been also simple: it is a number
of approximated mismatches divided by $m$ (i.e.\ the length of the
pattern). Unfortunately, we cannot design such a mapping. However,
we can design a set of mappings such that for a random mapping,
this condition holds with high probability.


\section{Probabilistically Separating Hashing} \label{s:mapping}
 In this section, our goal is to construct a random
hash $\pi$ for a given $D$ such that \eqref{Mapping} holds with
good probability. The set of hash functions $\mathcal{H_D}$ called
$C$-\emph{Probabilistically Separating Hashing }if it admits next
two conditions:
\begin{enumerate}
\item If the distance between $x,y$ is greater than $D$, then they
their hashing is different  i.e.\

  $\dist(x,y)\geq D \Rightarrow
  \forall \pi \in \mathcal{H_D}\ \pi(x)\neq \pi(y)$.
\item  $\forall x,y\in \Sigma,\ \Pr_{\pi \in \mathcal{H_D} }(\pi(x)\neq \pi(y)) \leq
  C\frac{\dist(x,y)}{D}$.
\end{enumerate}

Bartal at ~\cite{Bar96} gave a construction of $\log
\abs{\Sigma}$-{Probabilistically Separating Hashing } for finite
metrics after it was extended for graphs embedded in real normed
spaces at ~\cite{CCGGP98}. In section ~\ref{explicit} we will give
a simple construction for the case when alphabet is normed space
$\mathbb{R}^d$ with small $d$.

Notice that we only need to build such a hashing only once for
every alphabet. Therefore it can be done as a preprocessing
measure.


We are able to use $\pi_D$ in order to sample a subset of indices
for which the distance is not too small. We will now show that this
will also allow us to sample $X_D$ as we desire.

\begin{lem} \label{map}
Fix an offset $i$. Let $A=\{j\mid \pi_D(t_{i+j})\neq \pi_D(p_j)) \}$
be the set of mismatches under $\pi_D$ and $B=\{j\mid
D<\dist(t_{i+j},p_j)\leq 2D \}$ be the set of indices we are really
interested in sampling from them. Then:
\begin{enumerate}
\item $\bfE_{\pi_D\in \mathcal{H_D}}(\abs{A})=\calO(\frac{S\cdot C}{D})$ (where the
expectation is over the choice of $\pi_D$)
\item $B\subseteq A$ and $\frac{S_D}{D}\leq \abs{B} \leq \frac{2S_D}{D}$.
\end{enumerate}
Where $S=\sum_{j=0}^{m-1}\dist(p_j,t_{i+j})$ and $S_D=\sum_{j\in
B}\dist(p_j,t_{i+j})$.
\end{lem}
\begin{proof}
By linearity of expectation:
$$\bfE(\abs{A})=\sum_j\Pr(\pi_D(t_{i+j})\neq \pi_D(p_j)).$$
By definition of Probabilistically Separating Hashing we have:
\begin{equation*}
\sum_j\Pr(\pi_D(t_{i+j})\neq \pi_D(p_j))\leq \sum_j
\frac{C\dist(p_j,t_{i+j})}{D}=\frac{C\cdot S}{D}.
\end{equation*}
 This proves (1).

$B\subseteq A$ follows from theorem ~\ref{emmbed} and
$\frac{S_D}{D}=\sum_{j\in B}\frac{\dist(p_j,t_{i+j})}{D}$ but
$1\leq\frac{\dist(p_j,t_{i+j})}{D}\leq 2$ for $j \in B$ so
$\frac{S_D}{D}\leq \abs{B} \leq \frac{2S_D}{D} $
\end{proof}

\section{The Algorithm}
At this point, we have all the tools necessary in order to describe
the algorithm. The algorithm is based upon the application of
sampling algorithm, described previously,
 to the $C$-probabilistically separating hash provided in Sect.~\ref{s:mapping}.
 As a preprocessing phase, we construct for the metric space $(\Sigma,\dist)$, samples of hashing $\pi_D \in  \mathcal{H_D}$ for $D=2^i$.
The preprocessing algorithm therefore gets a metric space
$(\Sigma,\dist)$, where $\Sigma$ is the alphabet and $\dist$ is
the metric on it, and produces the
$\calO(\frac{1}{varepsilon^2}\log{\abs{\Sigma} \log m})$ hash
functions $\pi_D$ chosen at random from $\mathcal{H_D}$.

The main (i.e.\ query) algorithm gets a text $T=t_0t_1\ldots
t_{n-1}$ and a pattern $P=p_0\ldots p_{m-1}$ over the alphabet
$\Sigma$. For a fixed offset $i$ the result will be in
$((1-\varepsilon)S_i,(1+\varepsilon)S_i)$ with probability
$1-\bfe^{-t}$. The output of the algorithm is an array $R[0 \ldotdot
n-m-1]$ where $R[i]$ is an $\varepsilon$-approximation to
$S[i]=\sum_{j=0}^{m-1}\dist(p_j,t_{i+j})$ i.e:
\begin{equation}
\forall i\  \Pr(\abs{P[i]-S[i]}\geq \varepsilon S[i])\leq \bfe^{-t}
\end{equation}
We will now outline the idea of the algorithm. We want to approximate
$$m\bfE(X)=\sum_D m\Pr(D\leq X <2D)\bfE(X_D).$$
We will enumerate $D$, increasing it each time by a factor of 2,
and approximate $m\Pr(D\leq X <2D)\bfE(X_D)$. Fix some $D$ and
some offset $i$, let as before  $A=\{j\mid \pi_D(t_{i+j})\neq
\pi_D(p_j)) \}$, where $A$ depends on the random mapping $\pi_D$,
and $B=\{j\mid D<\dist(t_{i+j},p_j)\leq 2D \}$. Recall that $B
\subseteq A$, and that $\frac{\abs{B}}{ \abs{A}}$ is not too
small.

In order to approximate $\bfE(X_D)$ we will use the sampling
algorithm on $\pi_D(P)$ and $\pi_D(T)$. We get a random element in
$A$, and we check if this element is also in $B$. In order to
approximate $\bfE(X_D)$ we average the distances of elements found
in $B$.

In order to approximate $m\Pr(D\leq X <2D)=\abs{B}$ we use lemma
\ref{sample}. The probability that the sampling algorithm returns
``match'' is $q_0=\bfE(1-q)^{\abs{A}}$, and the probability that
it returns a mismatch from the set $B$ is
$q_1=\abs{B}q\bfE(1-q)^{\abs{A}-1}$. So,
$\abs{B}=\frac{q_1(1-q)}{qq_0}$.

Let's assume that we run the sampling algorithm $K$ times; then the total
number of matches is $m_0 \approx K q_0$ and the total number of
elements in $B$ is $m_1 \approx K q_1$. Let $M_1$ be an array of the
elements in $B$ which were found, including repetitions of elements
from $B$. $\abs{M_1} = m_1$.

We will approximate $\abs{B}$ by $\frac{m_1 (1-q)}{q m_0} $ because:
\be \label {1eq}
\abs{B} = \frac{\bfE(m_1) (1-q)}{q \bfE(m_0)}\ ,
\ee
 and approximate $\bfE(X_D)$ by $\frac{\sum_{j\in M_1}\dist(p_j,t_{i+j})
}{m_1}$. Therefore:
\begin{equation}
\begin{split}
&m\Pr(D\leq X <2D)\bfE(X_D) \approx\\
&\frac{(1-q)}{qm_0}\sum_{j\in M_1}\dist(p_j,t_{i+j})
\end{split}
\end{equation}

We will need to show that this approximation is narrow, i.e. that
the variance of the approximation is small. In order to do so, we
will need to choose $q$ s.t. $q \approx \frac{1}{\bfE{\abs{A}}}$.
In order to find such a $q$, we try a series of $q$'s, increasing
by a factor of 2 each time, and choose $q$ s.t. $m_0$ is large
enough and $q m_0$ is maximal. We prove that this produces a good
$q$ w.h.p.

We now write the complete algorithm. Set $K=\calO(\frac{1}{\varepsilon^2}
\cdot C\cdot t )$.

\algsetup{indent=2em}
\begin{algorithm}[H]
\caption{General distance algorithm} \label{alg:general_distance}
\begin{algorithmic}[1]
    \FOR {$D=B_{\dist}$ ;$D\geq 1$ ; $D=D/2$ }\label{FOR1}
       \FOR{$q=1/2$ ; $q>\frac{1}{m}$ ; $q=q/2$ }\label{FOR2}
        \FOR {$iter=1$; $iter \le K$; $iter=iter+1$} \label{FOR3}
            \STATE Choose a random $\pi \in  \mathcal{H_D}$
            \STATE run Sample($q, \pi(T),\pi(P)$). Save the result as the $iter$-th result for this $q$.
        \ENDFOR
    \ENDFOR
        \FORALL {offset in text $i$ }
            \STATE Calculate $m_0(i,q)$ for all $q$'s - the number
            of matches
            \STATE Among all $q$ such that $m_0\geq \bfe^{-4}\cdot K$ choose $q(i)$ s.t.\ $q(i) m_0(i,q(i))$ is maximal

                    \STATE Set $M_1$ to be the set of distances between $D$ and $2D$ for this $q$.
        \STATE Calculate
        $$S_D(i)=\frac{(1-q)}{q m_0}{\sum_{j\in M_1}\dist(t_{i+j},p_j)}$$

    \ENDFOR
  \ENDFOR
  \STATE for every offset $i$ return $R(i)=\sum_D S_D(i)$
\end{algorithmic}
\end{algorithm}

The running time of this algorithm is:
$\calO(\frac{1}{\varepsilon^2}n\log^2{m}\log{\abs{\Sigma}}\log{B_{\dist}})$.
This is because the running time is mostly dominated by the
Sampling function, which takes $\calO(n\log{m})$ time. The
Sampling function is executed
$\calO(\frac{1}{\varepsilon^2}\log{\abs{\Sigma}}
\log{m}\log{B_{\dist}})$ times.
\begin{thm}
For every offset $i$
\be
\Pr\left(\abs{R(i)-\sum_{j=0}^{m-1}\dist(p_j,t_{i+j})}\geq \varepsilon R(i)\right) \leq \bfe^{-t}\ ,
\ee
or in other words our algorithm returns $\varepsilon$-approximation w.h.p.
\end{thm}
The proof of this theorem appears in the appendix.

\section{Explicit hashing constructions for normed spaces}{\label{explicit}}
Now in this section let us construct explicit
$d$-Probabilistically Separating Hashing for normed space
$\mathcal{R}^d$ with ${\cal{L}}_p, 1 \leq p \leq \infty$ norm. The
main problem with previous constructions ~\cite{CCGGP98} is that
this hashing can't be calculated efficiently and usually it takes
$\calO(n^2)$ time to calculate one hash function. In case that
points not given in advance this may be bottleneck of the
algorithm.
 An other reason why this construction important is:
  if we have  $d$-Probabilistically Separating Hashing for space $X$ and we have embedding $f:Y \mapsto X$ with distortion $c$
  then we can construct $cd$--Probabilistically Separating Hashing for space $Y$.
  The problem of embedding metric spaces to real normed spaces where deeply investigated.

Our construction is the same for every norm ${\cal{L}}_p$.
 Let $\vec{\varepsilon}$ be a vector
of $d $ independent random variables with uniform distribution on
$[0,1]$. Define:
\begin{equation}
  \pi_D(\vec{x})=\left\lfloor \frac{\vec{x}}{D}-
  \vec{\varepsilon}\right\rfloor =
   \left(\left\lfloor\frac{x_1}{D}-
  \varepsilon_1\right\rfloor
  ,\ldots,\left\lfloor\frac{x_d}{D}-
  \varepsilon_d\right\rfloor\right)\ .
\end{equation}

\begin{thm}\label{emmbed}
The above mapping $\pi_D$ satisfies the next properties:
\begin{enumerate}
\item If the distance between $x,y$ is greater than ${D}\cdot d^{1/p}$, then their mapping
  is different i.e.\
  $\forall x,y \in R^d, \ \|x-y\|_p\geq D \cdot d^{1/p} \Rightarrow
  \pi_D(x)\neq \pi_D(y)$.
\item  $\forall x,y\in R^d,\ \Pr(\pi_D(x)\neq \pi_D(y)) \leq
  \frac{d\cdot\|x-y\|_p}{d^{1/p}D}$.
\end{enumerate}
\end{thm}
\begin{proof} As follows:
\begin{enumerate}
\item This is  trivial:
\begin{gather*}
Dd^{1/p}\leq \|x-y\|_{p} \\
\Rightarrow\exists i \abs{x_i-y_i}\geq D   \\
\Rightarrow\forall  \stackrel{\rightarrow}{\varepsilon} \pi_D(x)
\neq \pi_D(y).
\end{gather*}
\item If $\dist(x,y)\geq \frac{D}{cd} $ then this inequality is trivial.
Therefore assume  $\dist(x,y)\leq \frac{D}{cd}$:
\begin{gather*}
\Pr\left(\pi_D(x)\neq \pi_D(y)\right)\leq\\
  \sum_{i=1}^d{\bf{Prob}}(\pi_D(x)_i\neq \pi_D(y)_i)\leq\\
\sum_{i=1}^d\cdot\Pr\left(\lfloor \frac{x_i}{D}-
\varepsilon_i\rfloor\neq \lfloor \frac{y_i}{D}-
  \varepsilon_i\rfloor\right)\leq \\
  \sum_{i=1}^d\frac{\abs{x_i-y_i}}{D}=\frac{\|\vec{x}-\vec{y}\|_1}{D}\leq \frac{d\|\vec{x}-\vec{y}\|_p}{d^{1/p}D}
\end{gather*}
\end{enumerate}
\end{proof}
We choose to use $\pi$ as our embedding, and notice $\pi$ is easy
to calculate, assuming we already have the $c$-embedding $\sigma$.
This calculation can be done in $\calO(d\abs{\Sigma})$ time.
\begin{rem}
Consider the set $\{ \pi_{D}(x) \mid x \in \Sigma  \}$. While each
of its members is a vector of length $d$, when comparing these
vectors we are only interested in checking equality. Therefore, in
order to save space, we can replace each vector with a unique
number in $\{1 , \ldots, \abs{\Sigma} \}$.
\end{rem}
\section{Conclusions}
We have presented the first non-trivial algorithm for the
approximation of a large class of distances between text and
pattern. We believe that the techniques we have presented here have a wide
range of applications. A further interesting open question is to
generalize these techniques to the case where the distance is not necessary a metric.

\bibliographystyle{plain}
\begin{small}
\bibliography{paper}
\end{small}
\appendix
\section{The proof of the algorithm}
\begin{rem}
Here w.h.p. mean with probability more then $1-\bfe^{-t}$
\end{rem}
We will now prove that the algorithm indeed approximated the
distances for each $i$ w.h.p. We will only sketch
the proof.
\begin{proof}(of Algorithm)
Fix some offset $i$. Then for every $D$ we set $B=\{j\mid
D<\dist(t_{i+j},p_j)\leq 2D \}$ and $A=\{j\mid \pi_D(t_{i+j})\neq
\pi_D(p_j)) \}$ two sets. Notice that $\abs{A}$ is a random variable.
\begin{claim}\label{1claim}
W.h.p.\ for every $D$ there exist $q(D)$ s.t.\ $m_0(q)\geq
\bfe^{-4}K$ and $q\cdot m_0 \geq \frac{\bfe^{-4}K}{\bfE(\abs{A})}$
\end{claim}
\begin{proof}
There exist $q$ s.t.\ $\frac{1}{\bfE(\abs{A})}\leq q \leq
\frac{2}{\bfE(\abs{A})}$. The probability of a match for this $q$ is
$q_0=\bfE(1-q)^{(\abs{A})}$ by Jensen's inequality
$\bfE(1-q)^{\abs{A}}\geq (1-q)^{\bfE(\abs{A})}\geq \bfe^{-3}$.
$m_0(q)$  have binomial distribution $B(q_0,K)$ and so w.h.p.
$m_0\geq \bfe^{-4}K$. For this $q$  it also holds that $q\cdot m_0
\geq \frac{\bfe^{-4}K}{\bfE(\abs{A})}$. So for the $q$ that the algorithm
chose also holds that $q\cdot m_0 \geq
\frac{\bfe^{-4}K}{\bfE(\abs{A})}$.
\end{proof}
\begin{claim}
There exist a constant $\tilde{m_0}=\bfE(m_0) =K\bfE(1-q)^{\abs{A}}$. Such
that.
\[(1-\varepsilon)m_0(D)\leq
\tilde{m_0}(D)\leq(1+\varepsilon)m_0(D)\] w.h.p.
\end{claim}
\begin{proof}
This follows from the Chernoff bound. This is because $m_0$ is binomially
distributed variable $B(K,p)$ with $p\geq \bfe^{-4}$.
\end{proof}
Lets $c_D=\frac{(1-q(D))D}{q(D)\tilde{m_0}(D)}$ be a constant and $\displaystyle \tilde{S}(i)=\sum c_D\sum_{j\in M_1(D)}
\frac{\dist(t_{i+j},p_j)}{D}$
\begin{claim}
$\tilde{S}(i)$ is close to $R(i)$ w.h.p.\ i.e.
\[
(1-\varepsilon)R(i)\leq \tilde{S}(i)\leq(1+\varepsilon)R(i)
\]
\end{claim}
\begin{proof}
We can represent $R(i)$ as:

 \[
 R(i)=\sum_D{\sum_{j\in
M_1(D)}\frac{(1-q(D))D}{q(D)m_0(D)}\cdot\frac{ \dist(t_{i+j},p_j)
}{D}}\ .\]
By the previous claim $m_0(D)$  is close to $\tilde{m_0}(D)$.
\end{proof}
\begin{claim}\label{expectation}
\[
\bfE(\tilde{S}(i))=\sum_{j=0}^{m-1} \dist(p_i,t_{i+j}) .
\]
\end{claim}
\begin{proof}
\begin{equation}
\begin{split}
&\bfE(\tilde{S}(i))=\sum \frac{c_D}{D}\bfE(\sum_{j\in
M_1(D)}\dist(t_{i+j},p_j))=\\
&\sum \frac{c_D}{D}\bfE\abs{M_1(D)}\bfE(X_D) ,
\end{split}
\end{equation}
 But we know that:
\begin{equation}
\begin{split}
&\frac{c_D\bfE(\abs{M_1(D)})}{D}=\frac{(1-q(D))\bfE(m_1(D))}{q(D)\tilde{m_0}(D)}=\\
&\frac{(1-q(D))\bfE(m_1(D))}{q(D)\bfE({m_0}(D))}\ .
\end{split}
\end{equation}

By \eqref{1eq} we have that: $\frac{c_D\bfE(\abs{M_1(D)})}{D}=\abs{B}=m\Pr(D\leq X_D <2D)\ .$
So we have that:
\begin{equation}
\begin{split}
 &\bfE(\tilde{S}(i))=\sum_D m\Pr(D\leq X_D <2D)\bfE(X_D)=\\
 &m\bfE(X)=\sum_{j=0}^{m-1} \dist(p_i,t_{i+j})\ .
\end{split}
\end{equation}
\end{proof}
\begin{claim}\label{bound}
There exists a universal constant  $C$ s.t.\ w.h.p.\ :
\[c_D\leq C \frac{\varepsilon^2}{t} \sum c_D\abs{M_1(D)}\]
holds w.h.p.
\end{claim}
\begin{proof}
The previous claim showed us that $S(i) \leq \sum 2c_D\abs{M_1(D)}$ because $ \frac{\dist(p_i,t_{i+j})}{D}\leq 2 $ for $j\in M_1(D)$. Therefore, it's enough to prove that $c_D\leq C \frac{\varepsilon^2}{t}S(i)$\ .
By ~\ref{1claim} we know that $q m_0\geq \frac{\bfe^{-4}K}{\bfE(\abs{A})}$. So we have:
\be
c_D=\frac{(1-q(D))D}{q(D)\tilde{m_0}(D)}\leq \frac{C \bfE(\abs{A})D}{K}\ .
\ee
By ~\ref{map} $\bfE(\abs{A})D=\calO(S\cdot c\cdot d)$. $K=\calO(\frac{1}{\varepsilon^2}
\cdot c\cdot d\cdot t )$ So
\be
\frac{C \bfE(\abs{A})D}{K}=\calO(\frac{\varepsilon^2}{t}S)
\ee
\end{proof}
We will state the
following lemma without proving it because it follows from the Chernoff bound:
\begin{lem}\label{e-bound0}
There exists a universal constant $C$ s.t. for every sequence of
independent random variables $X_1,X_2\ldots X_n$ with $1\leq X_i
\leq2$, and a sequence of positive constants $c_1,c_2, \ldots c_n$
s.t. $c_i < C \frac{\varepsilon^2}{t} \displaystyle\sum_{i=1}^{n}
c_i$. Then:
$$
\Pr\left(\abs{\sum_{i=1}^n c_i\cdot X_i-\bfE(\sum_{i=1}^n c_i\cdot
X_i)}\geq \varepsilon\bfE(\sum_{i=1}^n c_i\cdot X_i)\right)\leq
\bfe^{-t}
$$
\end{lem}
\begin{claim}
\be \Pr(\abs{\tilde{S}-\sum_{j=0}^{m-1} \dist(p_i,t_{i+j})} \geq
\varepsilon \tilde{S}) \leq \bfe^{-t}. \ee
\end{claim}
\begin{proof}
$\tilde{S}=\sum c_i \frac{X_D}{D}$ by definition $1\leq\frac{X_D}{D}\leq 2$, by ~\ref{expectation} $\bfE(\tilde{S})=\sum_{j=0}^{m-1} \dist(p_i,t_{i+j})$ by ~\ref{bound} follows that $c_i < C \frac{\varepsilon^2}{t} \displaystyle\sum_{i=1}^{n}
c_i$ and therefore ~\ref{e-bound0} proves the claim
\end{proof}

\end{proof} 

\end{document}